\documentclass[vecphys]{svmult}

\usepackage{makeidx}         
\usepackage{graphicx}        
\usepackage{multicol}        
\usepackage[bottom]{footmisc}
\usepackage{amssymb}

\makeindex             

\begin{document}

\title*{Ultra-high energy cosmic rays: from GeV to ZeV}
\author{Gustavo Medina Tanco}
\institute{Instituto de Ciencias Nucleares, UNAM \\
\&  Instituto Astron\^omico e Geof\'{\i}sico, USP \\
\texttt{gmtanco@gmail.com}}
\maketitle

\section{Introduction}\label{Section:Introduction}

Cosmic ray (CR) particles arrive at the top of the Earth's
atmosphere at a rate of around $10^{3}$ per square meter per
second. They are mostly ionized nuclei - about 90\% protons, 9\%
alpha particles traces of heavier nuclei and approximately 1\%
electrons.

CRs are characterized by their high energies: most cosmic rays are
relativistic, having kinetic energies comparable to or somewhat
greater than their rest masses. A very few of them have
ultra-relativistic energies extending beyond $l0^{20}$ eV (tens of
joules).

In this series of lectures, delivered at the 2005 Mexican School
of Astrophysics, an overview of the main experimental
characteristics of the CR flux and their astrophysical
significance is given. Particular emphasis is given to the upper
end of the CR energy spectrum.

Unfortunately, due to space limitations, only a fraction of the
original content of the lectures is included in the present
manuscript. In particular, the production mechanisms are not
included and the fundamental topic of anisotropies is only dealt
with in a very superficial way.

\section{Energy spectrum}\label{Section:Spectrum}

Thus, the cosmic ray energy spectrum extends, amazingly, for more
than eleven orders of magnitude. All along this vast energy span,
the spectrum follows a power law of index $\sim 2.7$. Therefore,
the CR flux decreases approximately 30 orders of magnitude from
$\sim 10^3$ $m^2$ sec$^{-1}$ at few GeV to $\sim 1$ km$^{-2}$ per
century at 100 EeV.

\begin{figure}[h]
\begin{center}
\includegraphics*[width=12cm,angle=0]{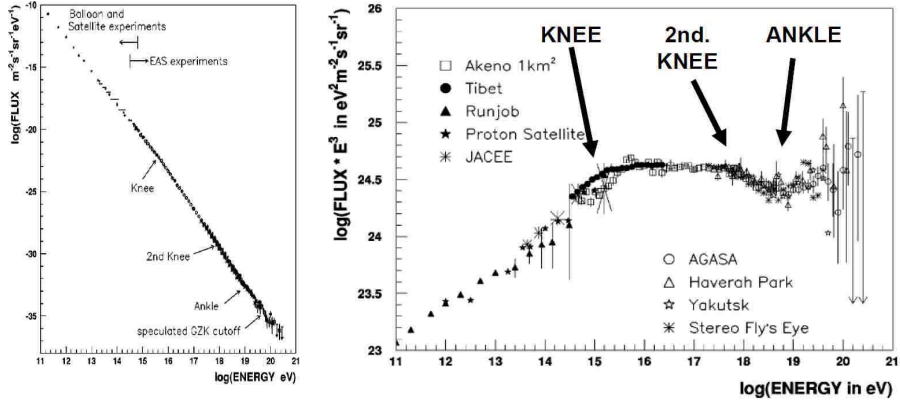}
\caption{ Cosmic ray energy spectrum and its main features: [left]
a remarkably uniform power law with [right] few bends knee (few
PeV), second knee (~0.5 EeV), ankle (EeV to few tens of EeV) and
the still poorly known highest energy tail. Adapted from
\cite{Nagano&Watson}} \label{fig:espectral_features}
\end{center}
\end{figure}

The only spectral features are a slight bending at around few PeV,
known as the first knee, another at approximately $0.5$ EeV known
as the second knee, and a dip extending from roughly the second
knee up to beyond $10$ EeV, known as the ankle (see, figure
\ref{fig:espectral_features}). Note that in the right panel the
spectrum is multiplied by $E^{3}$, a usual trick to highlight
features that otherwise would be almost completely hidden by the
rapidly falling flux.

The second knee has been observed in the vicinity of $4 \times
10^{17}$ eV by Akeno \cite{akeno_2ndknee}, Fly's Eye stereo
\cite{Bird1993,Bird1995,AbuZayyad2001}, Yakutsk
\cite{Yakutsk_2ndKAnklea,Yakutsk_2ndKAnkleb} and HiRes
\cite{HiRes2004}. The physical interpretation of this spectral
feature is uncertain at present. It may be either the end of the
Galactic cosmic ray component or the pile-up from pair creation
processes due to proton interactions with the cosmic microwave
background radiation during propagation in the intergalactic
medium.

The ankle, on the other hand, is a broader feature that has been
observed by Fly's Eye \cite{Bird1993,Bird1995,AbuZayyad2001}
around $3 \times 10^{18}$ eV as well as by Haverah Park
\cite{HP_Ankle} at approximately the same energy ($3 \times
10^{18}$ eV). These results have been confirmed by Yakutsk
\cite{Yakutsk_2ndKAnklea,Yakutsk_2ndKAnkleb} and HiRes
\cite{HiRes2004}. AGASA also observed the ankle, but they locate
it at a higher energy, around $10^{19}$ eV \cite{Takeda2003}. As
with the ankle, more than one physical interpretations are
possible, which are intimately related with the nature of the
second knee. The ankle may be the transition point between the
Galactic and extragalactic components, the result of pair creation
by protons in the cosmic microwave background, or the result of
diffusive propagation of extragalactic nuclei through cosmic
magnetic fields.

\subsubsection{Composition}\label{Section:Composition}

Certainly, one of the scientifically most relevant pieces of
information inside the transition energy interval previously
defined, is the precise chemical composition of the primary CR
flux as a function of energy.

Several techniques have been used to determine the composition of
cosmic rays along the spectrum and, in particular, in the highest
energy region \cite{Watson_composition}:  (i) depth of maximum of
the longitudinal distribution, $X_{max}$
\cite{Linsley_Xmax,Zha_Xmax}; (ii) fluctuations of $X_{max}$
\cite{fluctuations_Xmax_1,fluctuations_Xmax_2};(iii) muon density
\cite{AGASA_muons1,AGASA_muons2}; (iv) steepness of the lateral
distribution function \cite{steepness_Ave,steepness_Dova}; (v)
time profile of the signal, in particular rise time of the signal
\cite{risetime_Ave}; (vi) curvature radius of the shower front;
(vii) multi-parametric analysis, such as principal component
analysis and neural networks  \cite{multiparam_GMT}, etc..
Unfortunately, as is frequently the case in physics, whenever
several techniques are applied to measure the same physical
magnitude, correspondingly, several results are obtained and, not
always agreeable among themselves. As will be shown below, this is
critical to the understanding of the astrophysics of ultra high
energy cosmic rays.

In order to analyze the astrophysical implications of the
composition along the second knee/ankle energy region, it is more
instructive to start from much lower energies. A significant point
is the first knee. Following KASCADE \cite{kascade_results}, a
gradual change in composition is observed through the knee, from a
lighter to a heavier composition. The first knee is a broad
feature which can be understood as a composition of power law
energy spectra with breaks that are in agreement with a rigidity
scaling of the knee position.

\begin{figure}[h]
\begin{center}
\includegraphics*[width=10cm,angle=0]{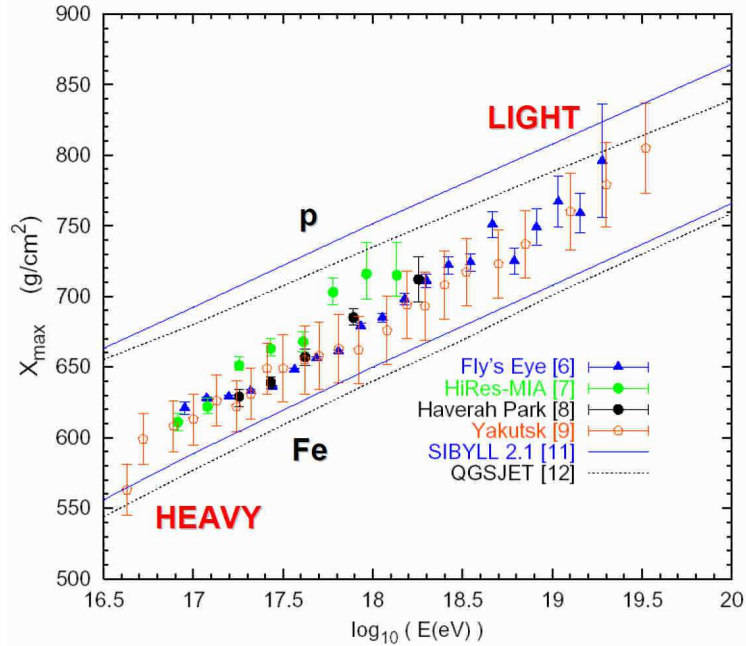}
\caption{Variation of $X_{max}$ with energy (elongation rate)
showing an apparent change in composition from heavy to light
nuclei from $\sim 30$ PeV to $\sim 30$ EeV. This variation is at
least possibly associated to the transition from Galactic to
extragalactic cosmic rays. The lines indicate theoretical
expectations corresponding to different hadronic interaction
models.} \label{fig:elongation_rate}
\end{center}
\end{figure}

Therefore, at energies above few times $10^{16}$ eV, the flux is
dominated by iron nuclei. These particles are of Galactic origin
and what is being detected is, very likely, the end of the
efficiency of supernova remnant shock waves as accelerators as the
Larmor radii or characteristic diffusion scale lengths of the
nuclei become comparable to the curvature radius of the remnants,
breaking down the diffusive approximation. If there are not more
powerful accelerators in the Galaxy, the Galactic cosmic ray flux
continue dominated exclusively by iron above $10^{17}$ eV up to
the highest energies produced inside the Milky Way. It must be
noted that, even if the previous results are quantitatively
dependent on the hadronic interaction model used in the data
analysis, they are qualitatively solid and there is considerable
degree of consensus on the existence of a progressive transition
in composition through the knee. At higher energies, the
composition has been measured by several experiments in the past,
e.g., Haverah Park, Yakutsk, Fly´s Eye, HiRes-MIA prototype and
HiRes in stereo mode (see figure \ref{fig:elongation_rate}).

The $X_{max}$ data suggest that, above $10^{16.6}$ eV, the
composition changes once more progressively from heavy to light.
At the lower limit of this energy interval, the composition is
still heavy, i.e., iron dominated, in accordance to Kascade
results. Nevertheless, at energies of $10^{19}$ eV, even if still
showing signs of a contamination  by heavy elements, it is more
consistent with a flux dominated by lighter elements.

\begin{figure}[h]
\begin{center}
\includegraphics*[width=10cm,clip,angle=0]{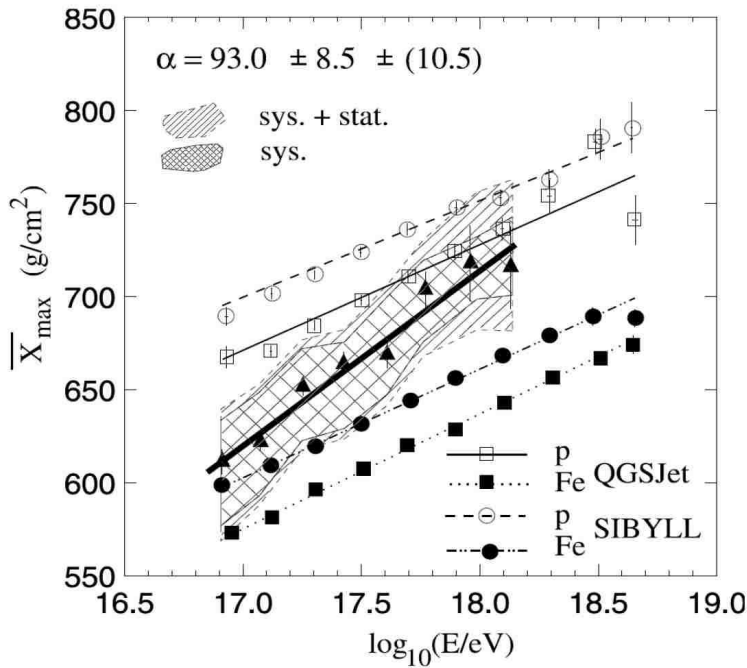}
\caption{Elongation rate measured by the HiRes-MIA prototype
showing a rapid change to a light composition above $10^{18}$ eV.}
\label{fig:elongation_rate_HiResMIA}
\end{center}
\end{figure}

Despite the fact that there is a consensus among most of the
experiments about the reality of this smooth transition, there is
no consensus about the rate and extent to which the transition
occurs. In fact, the combined data from the HiRes-MIA prototype
and HiRes in stereo mode, signal to a much more rapid rapid
transition from heavy to light composition (see figure
\ref{fig:elongation_rate_HiResMIA}), starting $10^{17}$ eV but
which would be over by $10^{18}$ eV \cite{HiResMIAXmax}. Beyond
that point, the composition would remain light and constant.

\begin{figure}[h]
\begin{center}
\includegraphics*[width=12cm,clip,angle=0]{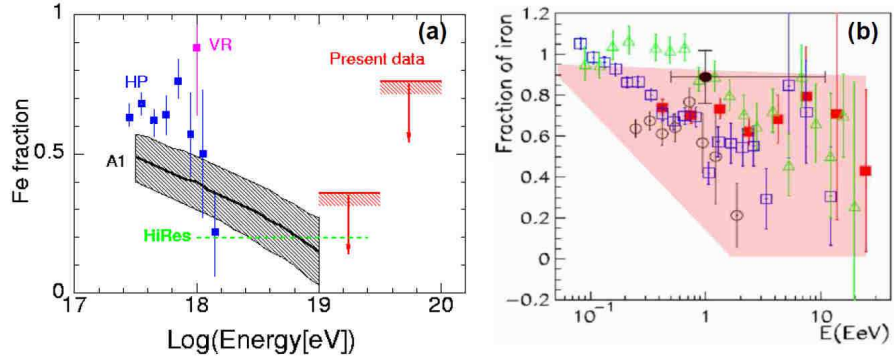}
\caption{(a) Variation of the iron fraction inside the transition
energy interval for various experiments. (b) Idem, highlighting
the uncertainties in composition inside the region encompassing
the second knee and the ankle: literally almost any abundance of
iron is allowed (adapted from \cite{steepness_Dova}).}
\label{fig:composition_confusion}
\end{center}
\end{figure}

The later scenario, however, is not supported by the data of other
experiments. Haverah Park, for example, shows a predominantly
heavy composition up to $10^{18}$ eV, followed by an abrupt
transition to lighter values compatible with HiRes stereo at
around $10^{18}$ eV (see, figure
\ref{fig:composition_confusion}.a). Volcano Ranch, even though
there is a single experimental point, is compatible with a heavy
composition still at $10^{18}$ eV, somewhat in accordance to
Haverah Park data. Akeno (A1), on the other hand, is consistent
with a continuation of the gradual transition from the second knee
all across the ankle up to at least $10^{19}$ eV, only reaching
there the same light composition that HiRes stereo claims from an
order of magnitude below in energy. It must also be noted that
above $10^{19}$ eV AGASA is only able to set upper limits for the
fraction of iron, but these limits are high enough to leave room
for much more complex astrophysical scenarios with a substantial
admixture of extragalactic ultra-high energy heavy elements
\cite{AGASA_composition}.

Figure \ref{fig:composition_confusion}.b shows a compendium of
several of the available measurements of composition between $\sim
10^{17}$ and $\sim 10^{19}$ eV, with their corresponding error
bars, under the simplistic assumption of a binary mixture of
protons and iron nuclei. The emerging picture is one complete
uncertainty, which has deep practical implications and imposes
severe limitations to theoretical efforts.

\begin{figure}[h]
\begin{center}
\includegraphics*[width=10cm,clip,trim=0 0 0 0]{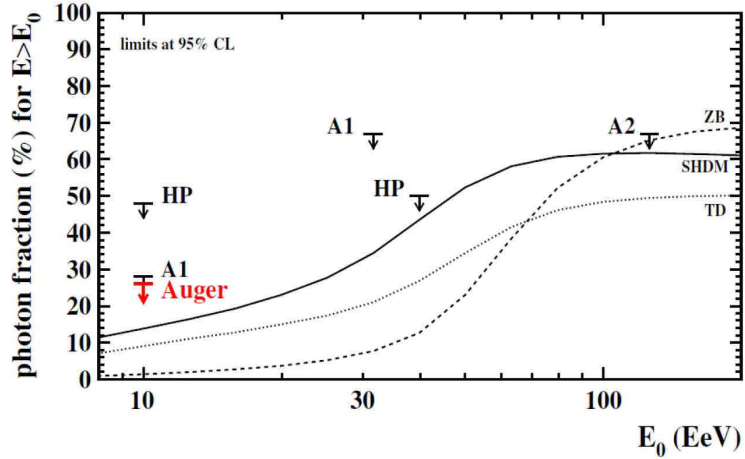}
\caption{ Upper limits (95\% CL) on cosmic-ray photon fraction for
Auger \cite{photon_limit_Auger}, AGASA (A1)
\cite{photon_limit_AGASA}, (A2) \cite{photon_limit_AGASA_Risse}
and Haverah Park (HP)
\cite{PhotonLimitHPAve2000,PhotonLimitHPAve2002} data compared to
some estimates based on TOP-DOWN models
\cite{PhotonLimitTOPDOWNestimates}. (Reproduced from
\cite{photon_limit_Auger}).} \label{fig:PhotonLimitAugerICRCPune}
\end{center}
\end{figure}

At energies beyond $\sim 10$ EeV the composition is essentially
unknown. However, it seems compatible with hadrons even if some
photon contribution cannot be discarded
\cite{photon_limit_AGASA,photon_limit_AGASA_Risse,photon_limit_Auger}
(see figure \ref{fig:PhotonLimitAugerICRCPune}). Although it is
implicitly regarded as purely protonic in many theoretical works,
only upper limit exists for the iron fraction (e.g.,
\ref{fig:composition_confusion}.b) and therefore not much can be
said until more accurate measurements are made available by new
generation experiments like Auger and TA.

\section{Galactic propagation}\label{Section:GalProp}

The fundamental question of cosmic ray physics is, "Where do they
come from?" and in particular, "How are they accelerated to such
high energies?". These are difficult questions not fully answered
after almost a century of history of the field.

Some very general hints can be obtained, however, through very
simple arguments. The interstellar and intergalactic mediae are
magnetized and, being charged, the CR are forced to interact with
these fields.

From the point of view of propagation of charged particles, the
Galaxy behaves as a magnetized volume, where the field is
structured on scales of kpc, with typical intensities of the order
of some few micro Gauss.

The Larmor radius of a nucleus of charge Ze can be conveniently
parameterized as:

\begin{equation}\label{Larmor_radius}
    r_{L,kpc} \approx \frac{1}{Z} \times \left( \frac{E_{EeV}}{B_{\mu G}} \right)
\end{equation}

\noindent where $E_{EeV}$ is the energy of the particle in units
of $10^{18}$ eV and $r_{L,kpc}$ is expressed in kpc.

Equation (\ref{Larmor_radius}) clearly shows that, given the
typical intensity of the magnetic fields present in the
interstellar medium (ISM), nuclei with energies below some few
tens of EeV, regardless of their charge, have Larmor radii much
smaller than the transversal dimensions of the magnetized Galactic
disk. They must, therefore, propagate diffusively inside the ISM.
This transforms the Galaxy in an efficient confinement region for
charged particles with energies below the second knee. The
confinement region is a flattened disk of approximately 20 kpc of
radius and thickness of the order of a few hundreds of pc.

Consequently, from the lowest energies and up to the second knee,
the Galaxy is undoubtedly the source of the cosmic ray particles
and of their kinetic energy. There is not a consensus about the
actual source of the particles in itself, but the two main lines
of thought propose either nuclei pre-accelerated at the
chromospheres of normal F and G type stars or ambient electrically
charged nuclei condensed in the dense winds of blue or red giant
stars \cite{RapporteurOGPune.Ptuskin,AccDustGrain.Meyer}. On the
other hand, several acceleration mechanism must be at play but it
is widely expected that the dominant one is first order Fermi
acceleration at the vicinity of supernova remnant shock waves.
Nevertheless, theoretically, the Galactic accelerators should
become inefficient between $\sim 10^{17}$ to $\sim 10^{18}$ eV.
This upper limit could be extended to $\sim 10^{19}$ eV if
additional mechanisms were operating in the Galaxy, e.g., spinning
inductors associated with compact objects or cataclysmic events
like acceleration of iron nuclei by young strongly magnetized
neutron stars through relativistic MHD winds
\cite{UHECRfromMHDrelatWinds}.

At energies above the second knee, particles start to be able to
travel from the nearest extragalactic sources in less than a
Hubble time. Consequently, at some point above $10^{17.5}$ eV a
sizable cosmic ray extragalactic component should be detectable
and become dominant above $10^{19}$ eV. Therefore, it is expected
that the cosmic ray flux detected between the second knee and the
ankle of the spectrum be a mixture of a Galactic and an
extragalactic flux, highlighting the astrophysical richness and
complexity of the region.

The type of propagation strongly depends on the charge of the
corresponding nucleus.

Protons with energies $\gtrsim 10^{17}$ eV have gyroradii
comparable or larger than the transversal dimensions of the
effective confinement region and, therefore, can easily escape
from the Galaxy. On the other end of the mass spectrum, just the
opposite occurs for iron nuclei that, even at energies of the
order of $10^{19}$ eV, have gyroradii $< 10^{2}$ pc and must be
effectively confined inside the magnetized interstellar medium.

The previous results are based only on the consideration of the
regular component of the Galactic magnetic field. However, there
exist a superimposed turbulent component whose intensity is at
least comparable to that of the regular field. Its spectrum seems
to of the Kolmogorov type, extending from the smallest scaled
probed, $\sim 10^0$ pc, to $L_{c} \sim 100$ pc, the correlation
length of the turbulent field.

Wave-particle interactions between cosmic rays  and MHD turbulence
are resonant for wavelengths of the order of the Larmor radius,
$\lambda \sim r_{L}$. This means that, for a nucleus of charge
$Z$, a critical energy can be defined,

\begin{equation}\label{critical_energy}
    r_{L,kpc} \approx \frac{1}{Z} \times \left( \frac{E_{EeV}}{B_{\mu G}}
    \right) \thickapprox L_{c}
     \; \; \; , \; \; \;
     L_{c} \approx 10^2 pc
     \; \; \; \Rightarrow \; \; \;
     E_{c,EeV} \approx 0.5 \times Z
\end{equation}

\noindent bellow which modes resonant with the particle gyroradius
exist that are able to efficiently scatter the particle in pitch
angle. Consequently, at energies bellow $E_{c}$ the diffusion
coefficient is small enough for the particle´s trajectory to be
diffusive. At energies above $E_{c}$, on the other hand, the
propagation is essentially ballistic.

Due to the interaction with the turbulent magnetic component,
protons experience a propagation regime very different than iron
nuclei inside the ankle energy region. Protons propagate
ballistically in the interstellar medium above $\sim 3 \times
10^{17}$ eV, while iron nuclei propagate diffusively even at
energies $\gtrsim 10^{19}$ eV.

Therefore, along the energy region extending from the second knee
up to almost the end of the ankle, all nuclei from p to Fe, i.e.
$1 < Z < 26$, experience a transition in their propagation regime
inside the interstellar medium changing gradually from diffusive
to ballistic as the energy increases.

\begin{figure}[h]
\begin{center}
\includegraphics*[width=11cm,angle=0]{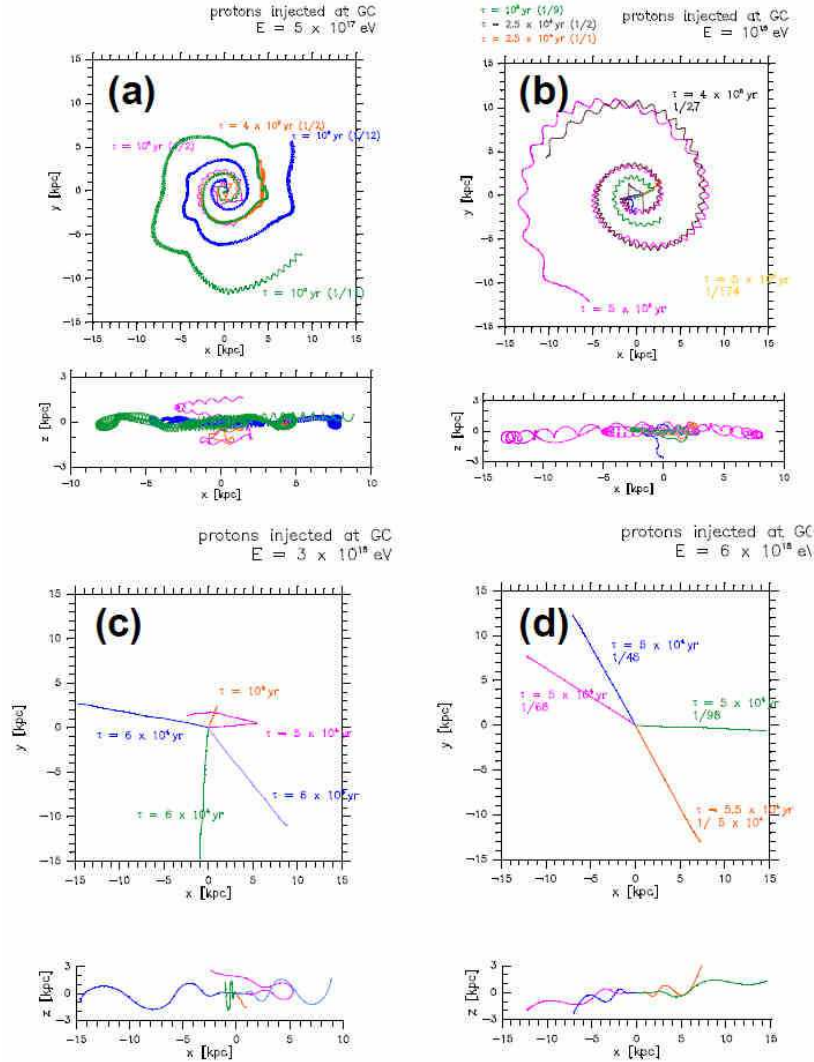}
\caption{Changes in propagation regime inside the Galaxy at
energies of the second knee and ankle.}
\label{fig:propagation_GalCenter}
\end{center}
\end{figure}

A pictorial example of how this transition takes place can be seen
in figure \ref{fig:propagation_GalCenter}.a-d
\cite{GalProp1EeV_GMT_Watson}, where it is shown how protons with
energies ranging from 0.5 to 6 EeV injected at the Galactic center
propagate out of the Galaxy assuming a characteristic BSS magnetic
model.

\begin{figure}[h]
\begin{center}
\includegraphics*[width=11cm,angle=0]{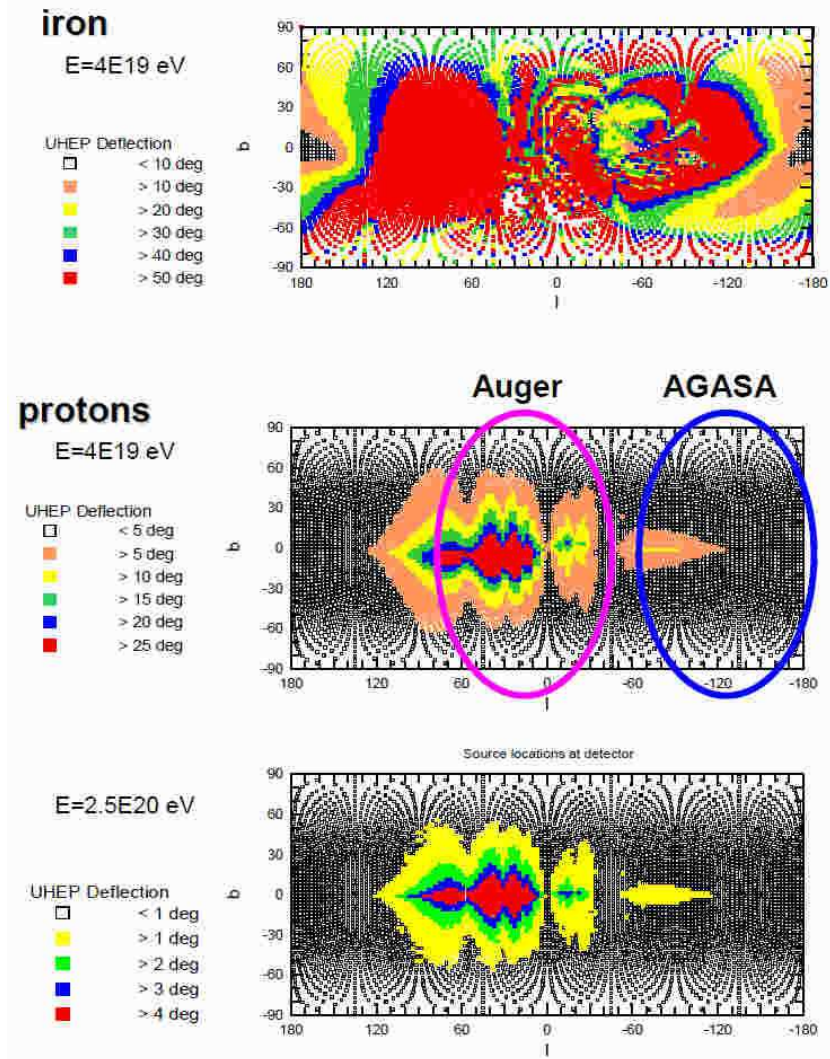}
\caption{Intrinsic deflections due to the GMF (BSSS model)
suffered by protons and Fe nuclei at $4\times 10^{19}$ eV and
protons at $2.5\times 10^{20}$ eV as a function of arrival
direction. Galactic coordinates are used.}
\label{fig:IntrinsicGalacticDeflections}
\end{center}
\end{figure}

It must be noted that, despite the fact that the deflections
induced by the Galactic magnetic field (GMF) diminish rapidly with
energy for all nuclei, they can still be important even at the
highest energies. This is more critical for heavier primaries and
for all nuclei traversing the central regions of the Galaxy.
Figure \ref{fig:IntrinsicGalacticDeflections} illustrated this
point by showing the intrinsic deflections experienced by proton
and iron nuclei as a function of arrival Galactic coordinates
\cite{IntrinsicGalacticDeflections_GMT}.

\begin{figure}[h]
\begin{center}
\includegraphics*[width=11cm,angle=0]{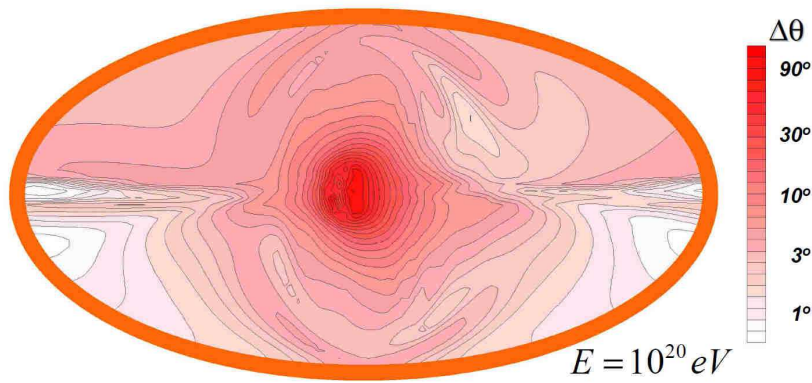}
\caption{Deflection suffered by 100 EeV protons due to a Han-type
GMF (see text) as a function of arrival direction. Galactic
coordinates are used in an Aitof projection.}
\label{fig:deflection100EeV_HanGMF}
\end{center}
\end{figure}

Figure \ref{fig:deflection100EeV_HanGMF} shows Galactic
deflections for 100 EeV protons for a more sophisticated GMF model
inspired on Han´s proposal \cite{HanGMF}. The GMF is modelled by a
disk BSSS component of 100 pc half thickness, embedded in an ASSA
halo and a dipolar component originated in a Southward magnetic
momentum anchored at the Galactic center. Galactic coordinates in
an Aitof projection are used.

\section{Extragalactic propagation}\label{Section:ExtragalProp}

\subsection{Superposition of the extragalactic and Galactic
fluxes}\label{Section:SuperpGalExtragal}

In the same way as the magnetic characteristic of the interstellar
medium allow Galactic particles at these energies to escape into
the extragalactic environment, extragalactic cosmic rays are also
able to penetrate inside the Galactic confinement region. But, of
course, extragalactic particles must first be able to reach us
from the nearest Galaxies in less than a Hubble time.

A crude approximation to this effect can be made in the following
way.

Faraday rotation measurements statistically impose to the
extragalactic magnetic field the following restriction
\cite{IGMF_Kronberg}:

\begin{equation}\label{faraday_constraint}
    B \times L_{c}^{1/2} \leqslant 1 nG \times Mpc^{1/2}
\end{equation}

\noindent where $L_{c}$ is the correlation length of the magnetic
field that we assume, somewhat arbitrarily, as being of the order
of $1$ Mpc. Assuming that the diffusion coefficient can be
estimated by the Bohm approximation,

\begin{equation}\label{Bhom_diffcoeff}
    K \thickapprox \frac{1}{3} r_{L} c
\end{equation}

\noindent and using equation (\ref{Larmor_radius}) the diffusion
coefficient can be written:

\begin{equation}\label{Bohm_diffcoeff2}
    K \thickapprox \frac{0.1}{Z} \left(  \frac{E_{EeV}}{B_{\mu G}} \right)
    \; \; \; \frac{\mbox{Mpc$^{2}$}}{\mbox{Myr}}
\end{equation}

The diffusive propagation time from an extragalactic source at a
distance $D$ can be estimated as:

\begin{equation}\label{diffusive_time}
    \tau \thickapprox \frac{D^{2}}{K}
\end{equation}

\noindent or, using equation (\ref{Bohm_diffcoeff2}):

\begin{equation}\label{diffusive_travel_time}
    \tau_{Myr} \thickapprox 10 \times D_{Mpc}^{2} \times Z \times
    \left(  \frac{B_{nG}}{E_{EeV}}  \right)
\end{equation}

\noindent Equation (\ref{diffusive_travel_time}) shows that there
is a rather restrictive magnetic horizon. Basically, no nucleus
with energy smaller than $10^{17}$ eV is able to arrive from
regions external to the local group ($D \sim 3$ Mpc). Taking as a
minimum characteristic distance $D = 10$ Mpc, which defines a very
localized region completely internal to the supergalactic plane
and even smaller than the distance to the nearby Virgo cluster,
only protons with $E > 2 \times 10^{17}$ eV, or Fe nuclei with $E
> 5 \times 10^{18}$ eV are able to reach the Galaxy in less than a
Hubble time.

Therefore, it is at the energies of the second knee and the ankle
that different nuclei start to arrive from the local universe.
Concomitantly, at these same energies, the magnetic shielding of
the Galaxy becomes permeable to these nuclei, allowing them to get
into the interstellar medium and, eventually, to reach the solar
system. Effectively, the energy interval from $\sim 2 \times
10^{17}$ to $10^{19}$ eV is the region of mixing between the
Galactic and extragalactic components of cosmic rays.

Above few times $10^{17}$ eV, the dominant interactions
experienced by cosmic rays are due to the cosmic microwave
background radiation (CMBR) and, additionally in the case of
nuclei, to the infrared background (CIBR) \cite{Berezinsky_book}.
The diffuse background in radio, despite its much lower density,
must in turn become important at high enough energies.

\begin{figure}[h]
\begin{center}
\includegraphics*[width=12cm,angle=0]{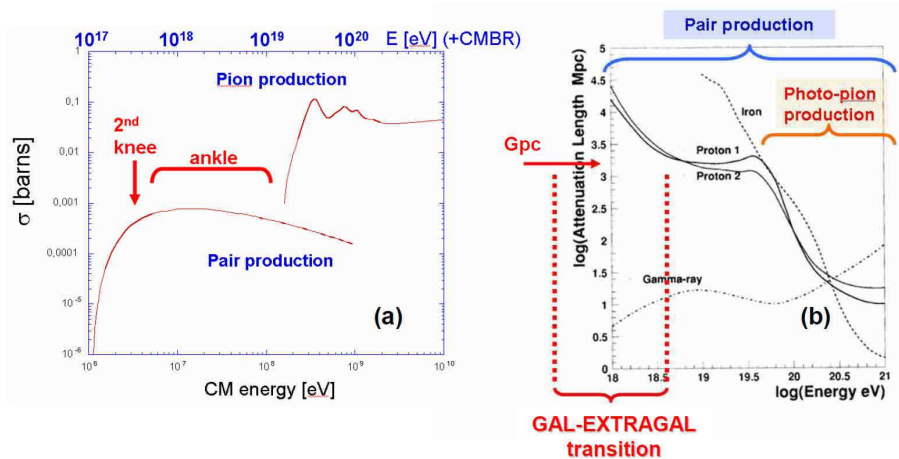}
\caption{(a) Cross section for pair production and pion production
in interaction with the CMBR. The positions of the second knee and
the ankle are also shown, demonstrating that electron positron
pair production is the relevant interaction in the
Galactic-extragalactic transition region. Note the similitude
between the shape of the cross section for this interaction and
the shape and location of the ankle. (b) Attenuation length in Mpc
as a function of energy \cite{Berezinsky_interactions}, showing
how the universe, which is opaque for at energies above the
photo-pion production threshold, becomes transparent to lower
energy barions.} \label{fig:interactions CMBR}
\end{center}
\end{figure}

At energies above $\sim 10^{19.2}$ eV, the dominant process is the
photo production of pions in interactions with the CMBR (see
figure \ref{fig:interactions CMBR}.a), which drastically reduces
the mean free path of protons to e few Mpc, making the universe
optically thick to to ultra-high energy cosmic rays
(\ref{fig:interactions CMBR}.b). This interaction, in the most
conservative models, should produce a strong depression in the
energy spectrum, with a major fall in the observed flux above
$10^{20}$ eV, the so called GZK cut-off
\cite{GZK_cut-off1,GZK_cut-off2}.

At energies smaller than $\sim 10^{19.2}$ eV, the dominant process
is the photo-production of electron-positron pairs in interactions
with the CMBR. At these lower energies the attenuation length
attain values of the order of Gpc and, therefore, the universe is
essentially optically thin to energetic baryons. CR observations
at these energies, sample the universe at cosmological distances,
contrary to the highest energies, that only sample a sphere of a
few tens of Mpc in diameter, a small portion of the local universe
\cite{Horizon_1,Horizon_2,Horizon_3}. Therefore, strictly from the
point of view of propagation in the extragalactic medium, in going
down from the highest energies to the transition region, the
observable horizon drastically increases from $10^{1}$ to $10^{3}$
Mpc, i.e., essentially the whole universe.

It can also be seen from figure \ref{fig:interactions CMBR}.a that
the dependence of the cross section with energy is suggestive,
since its shape resembles that of the ankle in the cosmic ray
energy spectrum. In fact, the structure of the ankle can be
explained exclusively as a result of pair photo-production by
nucleons travelling cosmological distances between the source and
the observer \cite{ankle_by_pairprod_Berezinsky}.

\begin{figure}[h]
\begin{center}
\includegraphics*[width=10cm,angle=0]{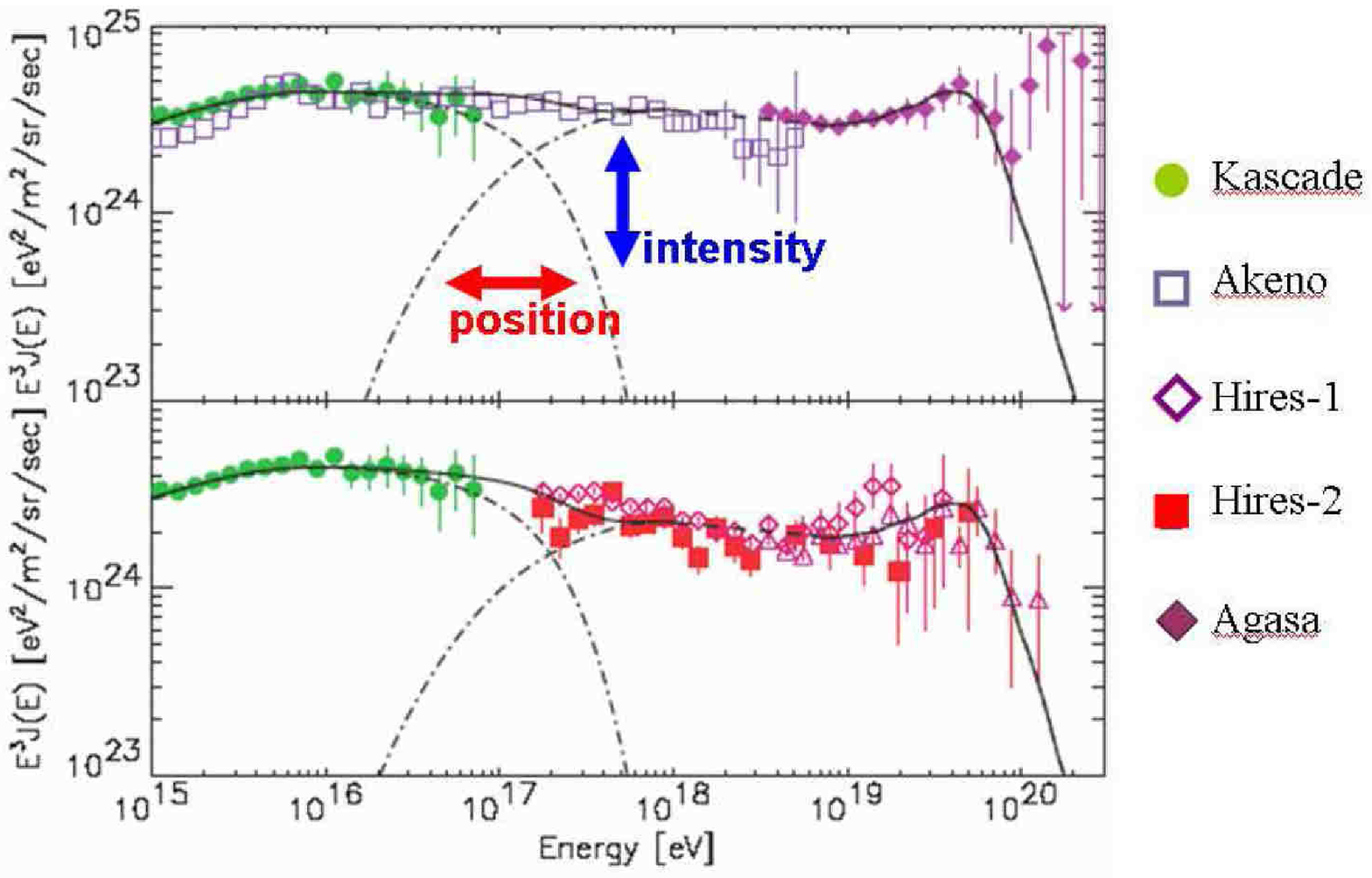}
\caption{Total cosmic ray spectrum from the combined data of
several experiments. From the theoretical point of view, the
transition region is highly complex and the Galactic and
extragalactic models suffer their most critical test since the
corresponding fluxes must be simultaneously matched both in
intensity and energy. (Adapted from:
\cite{spectrum_matching_lemoine}.)} \label{fig:spectrum_matching}
\end{center}
\end{figure}

The energy region where the superposition of the Galactic and
extragalactic spectra takes place is a theoretically challenging
region, where the smooth matching of the two rapidly varying
spectra has yet to be explained. It must be noted that, even if
the shape of the spectrum is important, it is by far insufficient
to decipher the the underlying astrophysical model. The Galactic
magnetic fields are intense enough to dilute any directional
information, which prevents the discrimination among the galactic
and extragalactic components from the arrival direction of the
incoming particles. The variation of the composition as a function
of energy turns then into the key to discriminate both fluxes and
to select among a variety of theoretical options.

As in the case of the interstellar medium, it is expected that the
intergalactic medium has a strong magnetic turbulent component
which can severely affect propagation
\cite{GMT_IGMFpropagation,GMT_IGM_nondiffusive_prop,GMT_ghosts,GMT_UHECRisotropic,GMT_IGMpropag_balistic,GMT_IGMpropag_voids}.
The correlation length estimated from Faraday rotation
measurements, $L_{c}$, is consistent with a maximum wavelength for
the MHD turbulence determined by the largest kinetic energy
injection scales in the intergalactic medium, $L_{max} \sim L_{c}
\sim 1$ Mpc. Therefore, analogous to equation
(\ref{critical_energy}):

\begin{equation}\label{critical_energy_IGM}
    r_{L,kpc} \approx \frac{1}{Z} \times \left( \frac{E_{EeV}}{B_{nG}}
    \right) \thickapprox L_{max}
     \; \; \; , \; \; \;
     L_{max} \approx 1 Mpc
     \; \; \; \Rightarrow \; \; \;
     E_{c,EeV} \approx 1.0 \times Z
\end{equation}

\begin{figure}[h]
\begin{center}
\includegraphics*[width=10cm,angle=0]{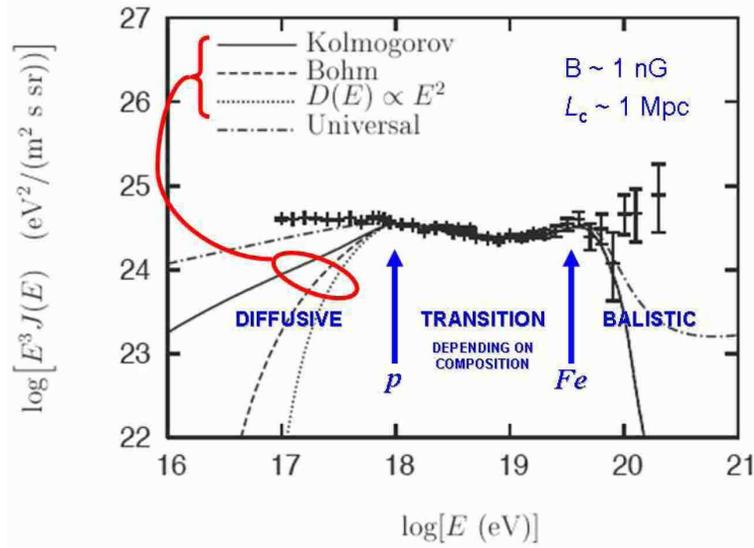}
\caption{ Correlation between the detailed structure of the lower
end of the extragalactic spectrum and the type of turbulence
present in the intergalactic medium. This emphasize the importance
of the intergalactic turbulent component in the observed matching
between the Galactic and extragalactic cosmic ray flux. (Adapted
from: \cite{IGM_LowEnd}.)}
\label{fig:IGM_propagation_KindOfTurbulence}
\end{center}
\end{figure}

For a given nucleus of charge $Ze$, the propagation is ballistic
for $E > E_{c}$ being diffusive otherwise. Therefore, protons are
ballistic above $\sim 10^{18}$ eV, but diffusive at the energies
of the second knee. Iron nuclei, on the other hand, propagate
diffusively along the ankle and even at energies as high as $\sim
5 \times 10^{19}$ eV. The boundaries for the transition between
the ballistic and diffusive propagation regime for proton and Fe
nuclei are shown in the figure
\ref{fig:IGM_propagation_KindOfTurbulence}.

Furthermore, besides the total intensity and the minimum
wavenumber, also the energy distribution among the different
modes, that is the type of turbulence present in the intergalactic
medium, have observational expression. In this case, the affected
portion of the extragalactic spectrum is the region of lower
energies, where the flux is strongly suppressed by magnetic
horizon effects. Figure \ref{fig:IGM_propagation_KindOfTurbulence}
shows clearly this effect for three different assumptions for the
diffusion coefficient.

\begin{figure}[h]
\begin{center}
\includegraphics*[width=11cm,angle=0]{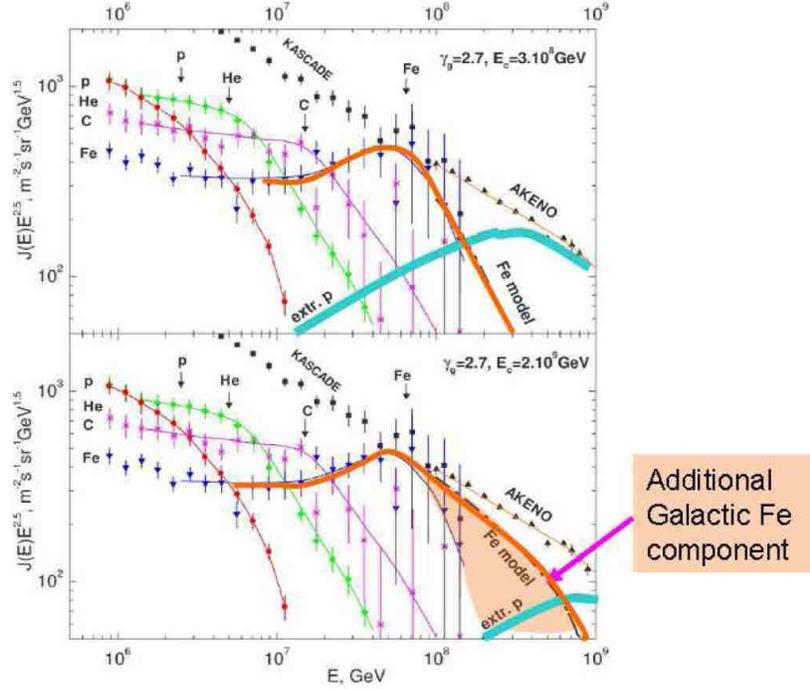}
\caption{ Impact of the detailed characteristics of the
extragalactic spectrum on our comprehension of the most powerful
acceleration mechanisms in our Galaxy. (Adapted from:
\cite{extraGal_Gal_matching}.)} \label{fig:matching_spectra_ankle}
\end{center}
\end{figure}

Obviously, this has profound theoretical implications not only for
the structure of the extragalactic magnetized medium, but also for
cosmic ray acceleration conditions inside the Galaxy. This is
exemplified in figure \ref{fig:matching_spectra_ankle} where it is
graphically illustrated that, by subtracting a given extragalactic
spectrum from the observed total spectrum, conclusions can be
drawn about relevant aspects of the Galactic component. For
example, an extragalactic spectrum that has a small contribution
at low energies, can imply the existence of additional
acceleration mechanisms in the Galaxy other than the shock waves
of supernova remnants.

\subsection{The highest energies}\label{Section:PropagationHighestEnergies}

\subsubsection{Propagation of protons}\label{Section:ProtonProp}

As was mentioned in the previous section
\ref{Section:SuperpGalExtragal} (see figure \ref{fig:interactions
CMBR}), above the threshold for photo-pion production by protons
interactiing with the CMBR, $\sim 40$ EeV the universe becomes
rapidly opaque for hadrons as the attenuation length goes down to
values as low as $\sim 10$ Mpc at few $\times 10^{20}$ eV. This
determines a relatively small (in cosmological terms) maximum
distance scale, $R_{GZK}\backsimeq 50$--$100$ Mpc, to the sources
that are able to contribute appreciably to the detected CR flux.

Under very general assumptions regarding the nature of the
primaries and the cosmological distribution of the sources,
photo-pion production should lead to the formation of a pile-up
immediately followed by a severe reduction in CR flux, popularly
known as the GZK cut-off. The existence of this spectral feature
was proposed short time after the discovery of the CMBR
\cite{GZK_cut-off1,GZK_cut-off2} but its actual existence is still
a matter of considerable debate.

At present, there are conflicting measurements coming from two
different experiments: AGASA and HiRes. The first one is a surface
detector while the second one is a fluorescence detector, which
further complicates the comparison of their results.

\begin{figure}[h]
\begin{center}
\includegraphics*[width=11cm,angle=0]{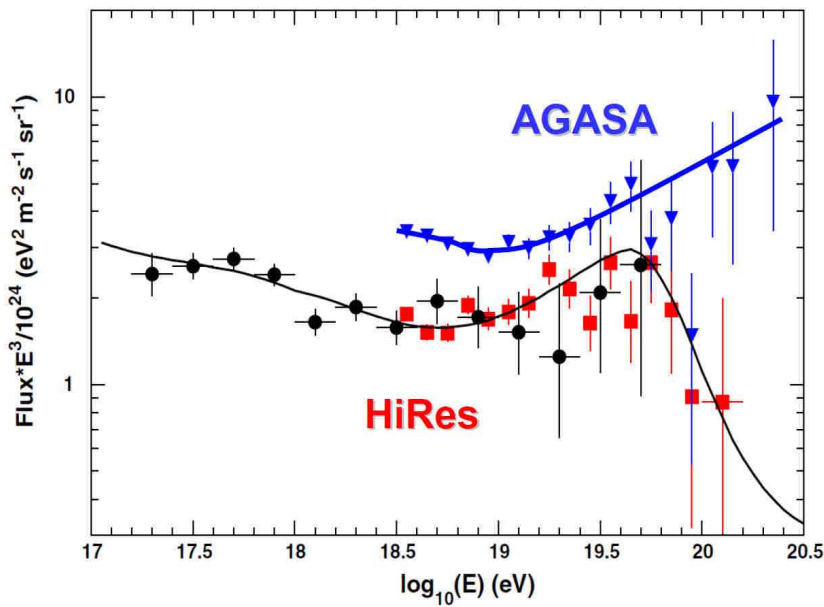}
\caption{Comparison between the AGASA and Hires monocular spectra.
(Adapted from: \cite{HiResMonoSpectrum}.)}
\label{fig:spectrumAGASAHiRes}
\end{center}
\end{figure}

As shown in figure \ref{fig:spectrumAGASAHiRes}, the differences
are not only quantitative but, fundamentally, qualitative. While
HiRes apparently shows the expected GZK flux suppression, AGASA
seems incompatible with this result, showing an energy spectrum
that extends undisturbed well beyond $100$ EeV. There is also an
apparently large difference in flux between both experiments at
energies below the cut-off. However, the fact that the energy
spectrum has been multiplied by $E^3$ in figure
\ref{fig:spectrumAGASAHiRes} should be taken into account when
assessing the significance of such difference.

Actually, both results might be reconciled at the $1.5 \, \sigma$
level by re-scaling the energy of either of the experiments by
30\% or by 15 \% each
\cite{SignificanceHiResAGASADifferenceDeMarco}. The Auger
Observatory, being the largest detector ever built and having
hybrid capacity (i.e., simultaneous fluorescence and surface
detection \cite{AugerNIM}) has the potential, but not enough
statistics at present, to give a definite answer to this
fundamental problem \cite{AugerICRCspectrum}.

The absence of the GZK cut-off, if confirmed, could be compatible
with a wide range of astrophysical scenarios. At least three
possibilities can be considered, some rather conservative, some
more exotic:

\begin{itemize}

\item The distance scale between sources could be large enough
that, by chance, the few (or single) sources inside the GZK sphere
dominate the flux, the rest of the population of UHECR
accelerators being too distant to contribute appreciably to the
observed flux at Earth.

\item The primary CR might be particles that do not interact with
photons or do so at much larger, and as yet unobserved, energies.
These could be familiar standard model particles that present
unexpected behavior at ultra-high energies, like neutrinos with
hadronic cross section that can develop showers in the atmosphere
resembling those expected from proton primaries
\cite{NeutrinosHadronicCrossSection1,NeutrinosHadronicCrossSection2}.
Another possibility could be a new stable hadron, heavier than a
nucleon, for which the threshold for photo-meson production would
be at higher energy. An example of the latter would be uhecrons,
e.g., a uds-gluino bound state \cite{UHECRONSChung}.

\item The primaries might be normal hadrons, but Lorentz
invariance, which has never been previously tested at $\gamma \sim
10^{11}$, could be violated at ultra-high energies, hampering
photo-meson production
\cite{LorentzInvViolationSato,LorentzInvViolationKirzhnits,LorentzInvViolationColemanGlashow}.
The small violation of Lorentz invariance required might be the
result of Planck scale effects
\cite{LIViolationPlanckScaleAloisio,LIViolationPlanckScaleAlfaro}.

\item The observed spectrum could be the superposition of two
components: (a) a hadronic component with a GZK cut-off and (b) a
harder, top-down component that becomes dominant above $\sim 100$
EeV. These second component could be originated in the decay or
annihilation of super heavy dark matter or topological defects
\cite{TopDownRevSigl,TopDownRevBerezinsky,TopDownRevKuzmin,TopDownSarkar}.
These scenarios have the general disadvantage of overproducing
ultra-high energy neutrinos and photons rather than nucleons,
which seems to be increasingly constrained by the observation
\cite{photon_limit_Auger,PhotonLimitAGASAYakutsk}. Nevertheless,
there could still be models, like those involving necklaces, that
could present an appreciable baryon content at energies $\gtrsim
100$ EeV \cite{TopDownSpectraAloisio}. In any case, these models
suffer from a discomforting level of fine tuning with respect to
the normalization of the intensities of the GZKed and the top-down
spectra.

\end{itemize}

It must also be noted that the presence of the GZK flux
suppression does not imply the non-existence of supra-GZK
particles in the CR flux. These have certainly been detected by at
least Volcano Ranch \cite{VRUHECRLinsley1,VRUHECRLinsley2}, Fly´s
Eye \cite{FEUHECR3e20}, AGASA, HiRes and, more recently, Auger
\cite{AugerICRCMantsch}. This means that, the detection of the GZK
feature does not solve the puzzle about the generation of UHECR.

\subsubsection{Propagation of UHE photons}\label{Section:PhotonProp}

The propagation of photons is dominated by their interaction with
the photon background. The main processes are photon absorption by
pair-production on background photons ($\gamma \gamma_b
\rightarrow e^+ e^-$), and inverse Compton scattering of the
resultant electrons on the background photons. These two processes
acting in a chain are responsible for the rapid development of
electromagnetic cascades in the intergalactic or interstellar
media, draining energy to the sub-TeV region.

For a given UHE photon of energy $E_{\gamma}$, the minimum
background photon energy, $E_{b}$, for electron-positron pair
production is:

\begin{equation}\label{eq:PPthreshold}
E_{b} = \frac{m_{e}^{2}}{E_{\gamma}} \simeq \frac{2.6 \times
10^{11}}{E_{\gamma,eV}} \,\,\, \mbox{eV}
\end{equation}

\noindent and the corresponding cross section peaks near the
threshold: $\sigma_{PP} \propto (m_{e}^{2}/s)*ln(s/2m_{e})$.
Inverse Compton scattering, on the other hand, has no threshold
but its cross section is also largest near the $\gamma \gamma_{b}$
pair production threshold. Therefore, the most efficient
background for both processes is given by equation
(\ref{eq:PPthreshold}). For UHE this means that the cosmic radio
background, whose magnitude is highly uncertain, is dominant
followed by the CMBR below $10^{17}$ - $10^{18}$ eV. At
progressively lower energies, the CIRB and optical background are
important.

In the Klein-Nishina limit, $s \gg m_{e}^2$, one of the components
of the $\gamma \gamma_{b}$-produced pair carries most of the
energy of the energy of the UHE photon. This leading particle,
afterwards, undergoes Compton scattering in the same limit, for
which the inelasticity is very near $1$. Therefore, the Compton
up-scattered photon still has an appreciable fraction of the
energy of the original UHE photon. The presence of magnetic fields
in the medium may speed up the development of the cascade by
draining the electron and positron energy due to synchrotron
radiation. The larger the fields, the smaller the penetration.

\begin{figure}[h]
\begin{center}
\includegraphics*[width=11cm,angle=0]{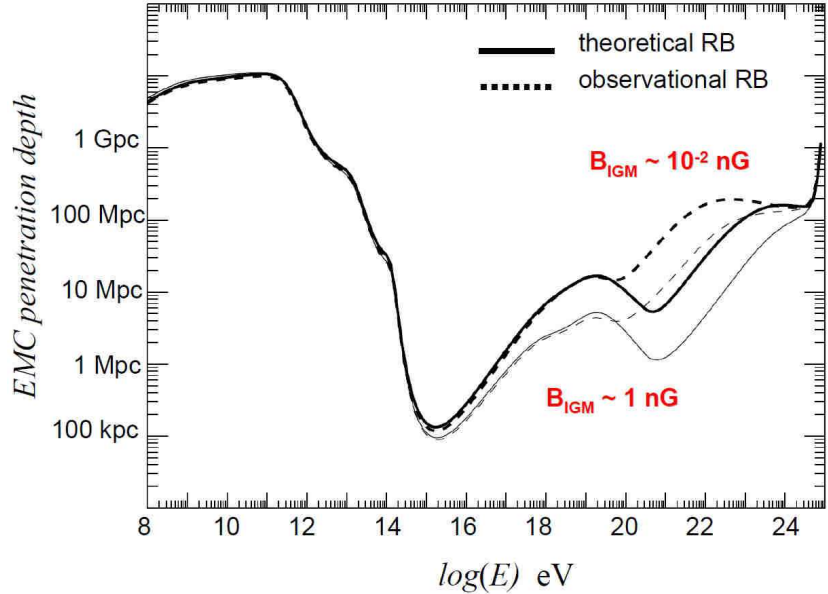}
\caption{Penetration of electromagnetic cascades in the
intergalactic medium. Solid lines correspond to a fiduciary value
of 1 nG for the IGMF and dotted lines to a very low IGMF,
$10^{-2}$ nG. Thick and thin lines correspond to different
estimates of the cosmic radio background. (Adapted from:
\cite{TopDownRevSigl}.)} \label{fig:PhotonPropag}
\end{center}
\end{figure}

The electromagnetic cascades produced in this way can propagate an
effective distance that is much larger than the interaction length
yet, severely limit our UHE-$\gamma$ horizon to the nearest
regions of the supergalactic plane. Figure \ref{fig:PhotonPropag}
\cite{TopDownRevSigl} shows the effective penetration length of
electromagnetic cascades for two different estimates of the cosmic
radio background and two different average intergalactic magnetic
field intensities.

Since the single pair cross section decreases as $\ln(s) / s$ for
$s \gg m_{e}^{2}$, multiple pair production becomes important at
extreme energies. Thus, double pair production ($\gamma \gamma_b
\rightarrow e^+ e^-e^+ e^-$) begins to dominate above $\sim
10^{21}$ -- $10^{23}$ eV. The relevant process for electrons is
triple pair production ($e^{\pm} \gamma_b \rightarrow
e^{\pm}e^-e^+$), whose attenuation becomes dominant at $\sim
10^{22}$ eV. Other processes (e.g., moun, tau or pion pair
production, double Compton scattering, gamma scattering and pair
production of single photons in magnetic fields) are in general
negligible for electromagnetic cascade development. However, at
energies in excess of $10^{24}$ eV, the pair production of single
photons in the Galactic magnetic field should eliminate all the
photons above that energy from specific lines of sight, generating
an arrival direction anisotropy.

The penetration lengths shown in figure \ref{fig:PhotonPropag},
combined with the threshold energies given by equation
(\ref{eq:PPthreshold}), imply that the injection of UHE-$\gamma$
in the intergalactic medium results in the pile-up of photons at
energies below 100 MeV, whose contribution to the diffuse cosmic
$\gamma$ background is already observationally constrained by
EGRET. This overproduction of low energy diffuse photons is a
strong restriction for top-down UHECR production models.

Besides the limitations imposed by the possible overproduction of
low energy photons, the results in figure \ref{fig:PhotonPropag}
have other profound implications for top-down scenarios. Models
that claim decay or annihilation of DM, in general, tend to
produce mainly photons and only e few percent baryons. Therefore,
in such models, most of the detected CR should be photons from our
own Galactic halo, with perhaps some localized contribution from
Andromeda. The products of the decay of more distant dark matter
would be cleared from photons, and only the small fraction of
remaining baryons, suppressed by GZK effects, would give a
positive contribution at Earth. In any case, for most top-down
models, photons should be dominant above the GZK cut-off, but with
perhaps a sizable baryonic component.

\subsubsection{Propagation of nuclei}\label{Section:NucleiProp}

Heavy nuclei are attenuated by basically to processes:
photodisintegration and electron-positron pair creation b y
interaction with background photons.

\begin{figure}[h]
\begin{center}
\includegraphics*[width=11cm,angle=0]{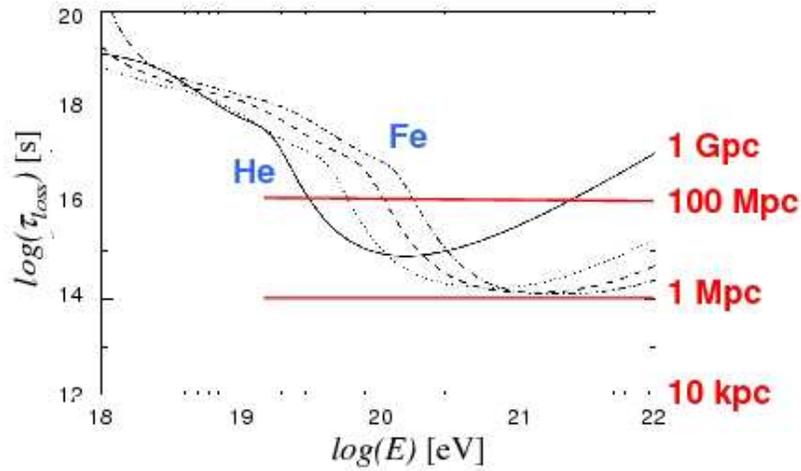}
\caption{Energy loss time (right axis: length) vs. energy for
photodisintegration on background photons: radio, CMB and IR.
Helium, Carbon, Silicon and Iron are shown. Single, double and
multi-nucleon emissions are included. (Adapted from:
\cite{NucAttenuationLengthBertone}.)} \label{fig:NucleiPropag}
\end{center}
\end{figure}

For the same total energy, the threshold photo-pion production for
a nucleus of mass A increases to $E_{th} \simeq 4 \times 10^{19}
\times A$. Therefore, given the energies observed at present, pion
production is not relevant for nuclei heavier than He.

Figure \ref{fig:NucleiPropag} shows that below $\sim 20$ EeV, all
nuclei are able to travel for virtually a Hubble time, while Fe
can do the same up to $\sim 100$ EeV. Above that energy, nuclei
start to disintegrate fast and the loss time are highly reduced.

It is clear from composition observations around the first knee of
the cosmic rays spectrum that the acceleration mechanism by shock
waves, either first order Fermi or drift are limited by the
magnitude of the radius of curvature of the shock. This results in
the preferential acceleration of large charge ($eZ$) nuclei, those
with the smallest Larmor radius, to the highest energies.

Even if the mechanism responsible for the acceleration of the
ultra-high energy extragalactic component is essentially unknown,
the most conservative view points to bottom-up mechanisms. If the
later is actually the case, then the most economic assumption is
that chock wave acceleration.

In this minimalistic, but still realistic, scenario the most
likely high energy output would be heavy nuclei, very likely Fe as
in the Galactic case. At lower energies, progressively lighter
nuclei should be observed due to two factors: (a) the acceleration
process in itself and (b) the photo-disintegration on flight of
the heavier nuclei due to their interaction with the CIBR. The
latter process is very efficient, and can extract approximately
one nucleon every few Mpc at the highest energies, depending on
the CIBR level. Since photodisintegration occurs, to a good
approximation, at constant energy per nucleon, disintegration of a
nucleus $A$ at energy $E$ will produce light nuclei at energies
$nE/A$ and $(A-n)E/A$ with preferentially small $n$ (e.g., $n=1,
2, 3, 4$).

\begin{figure}[h]
\begin{center}
\includegraphics*[width=12cm,angle=0]{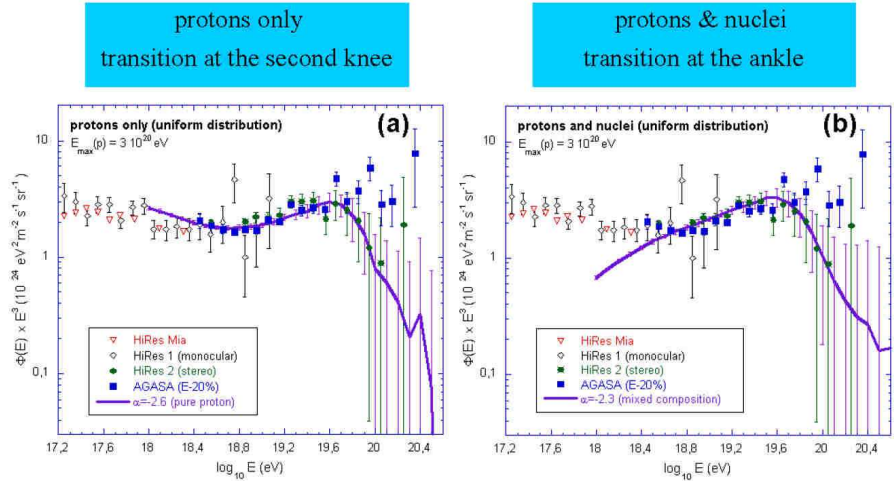}
\caption{Different primary compositions may produce extragalactic
spectra essentially indistinguishable at the highest energies.
However, as shown in (a) for protons and (b) for heavier mix, the
spectra are considerably different at lower energies below the
bottom of the ankle. (Adapted from:
\cite{composition_ankle_Allard}.)}
\label{fig:mixed_composition_effects1}
\end{center}
\end{figure}

Figure \ref{fig:mixed_composition_effects1} shows that power law
spectra injected at cosmological sources with different
compositions can produce experimentally very similar at the
highest energies. Nevertheless, they can always be distinguished
at smaller energies in the ankle region.

In particular, figure \ref{fig:mixed_composition_effects1}.a
showshow a purely protonic flux can reproduce the ankle feature
solely as an effect of photo production of electron-positron pairs
in interactions with the CMBR. In this case, the transition
between the Galactic and extragalactic fluxes must be located at
the second knee or very near to it.

Figure figure \ref{fig:mixed_composition_effects1}.b, on the other
hand, shows that, for a heavier mixed composition, the
extragalactic spectrum falls down steadily with decreasing energy.
In this scenario, the ankle must be the result of the composition
between the Galactic and extragalactic spectra. Moreover, the
composition will be a strong function of energy inside this
interval, giving an additional tool to assess details of the
astrophysical model.

The effects of photo-disintegration are clearly shown in figure
\cite{extraGal_iron_Yamamoto}, where the evolution of a pure iron
injected spectrum as it propagates out from the source. Blue
histograms correspond to p, white to the original surviving Fe and
red to intermediate mass nuclei.

As the distance to the source increases, intermediate nuclei are
produced at increasingly smaller masses and less total energy
(fragmentation takes place at roughly constant mass per nucleon).
At distances of the order of the GZK horizon, the region with
larger mixing of nuclei , i.e., with a larger composition
gradient, is the ankle. Consequently, this is the ideal region for
the discrimination of the primary composition from local
composition measurements.

\begin{figure}[h]
\begin{center}
\includegraphics*[width=9cm,angle=0]{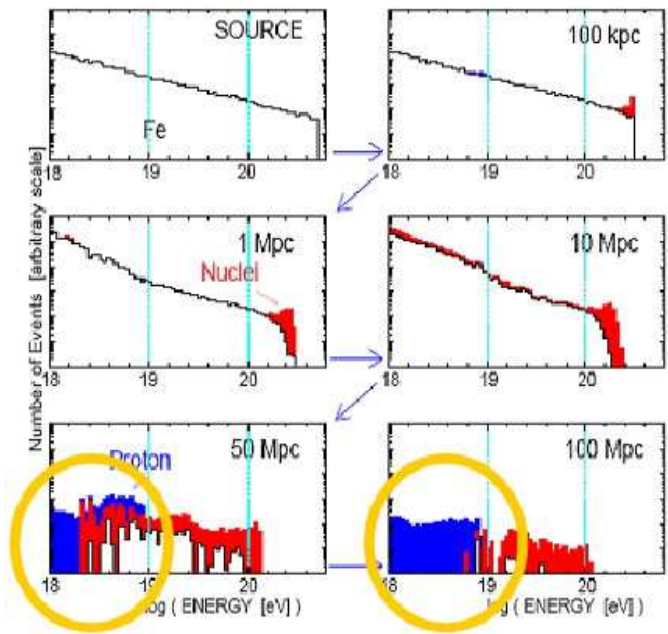}
\caption{ Variation of the composition as a function of distance
to the source for pure Fe power law injection. Blue histograms
correspond to p, white to the original surviving Fe and red to
intermediate mass nuclei. (Adapted from:
\cite{extraGal_iron_Yamamoto}.)}
\label{fig:extraGal_iron_injection}
\end{center}
\end{figure}

\section{Cosmic magnetic fields and anisotropy}\label{Section:Anisotropy}

Luminous matter, as traced by galaxies, as well as dark matter, as
traced by galaxies and clusters large scale velocity fields, is
distributed inhomogeneously in the universe. Groups, clusters,
superclusters, walls, filaments and voids are known to exist at
all observed distances and are very well mapped in the local
universe. Hence, the distribution of matter inside the GZK-sphere
is highly inhomogeneous and so is, very likely, the distribution
of UHECR sources.

Synchrotron emission and multi-wavelength radio polarization
measurements show that magnetic fields are widespread in the
Universe. But how do they encompass the structure seen in the
distribution of matter we do not yet know \cite{IGMF_Kronberg}.

The available limits on the IGMF come from rotation measurements
in clusters of galaxies and suggest that $B_{IGM} \times
L_{c}^{1/2} < 10^{-9}$ G $\times$ Mpc$^{1/2}$
\cite{IGMF_Kronberg}, where $L_{c}$ is the field reversal scale.
Note, however, that this kind of measurement doesn't set an actual
limit to the intensity of the magnetic field unless the reversal
scale is known along a particular line of sight. The latter means
that, depending on the structure of the IGMF, substantially
different scenarios can be envisioned that are able satisfy the
rotation measurement constraints.

\begin{figure}[h]
\begin{center}
\includegraphics[width=0.8\textwidth]{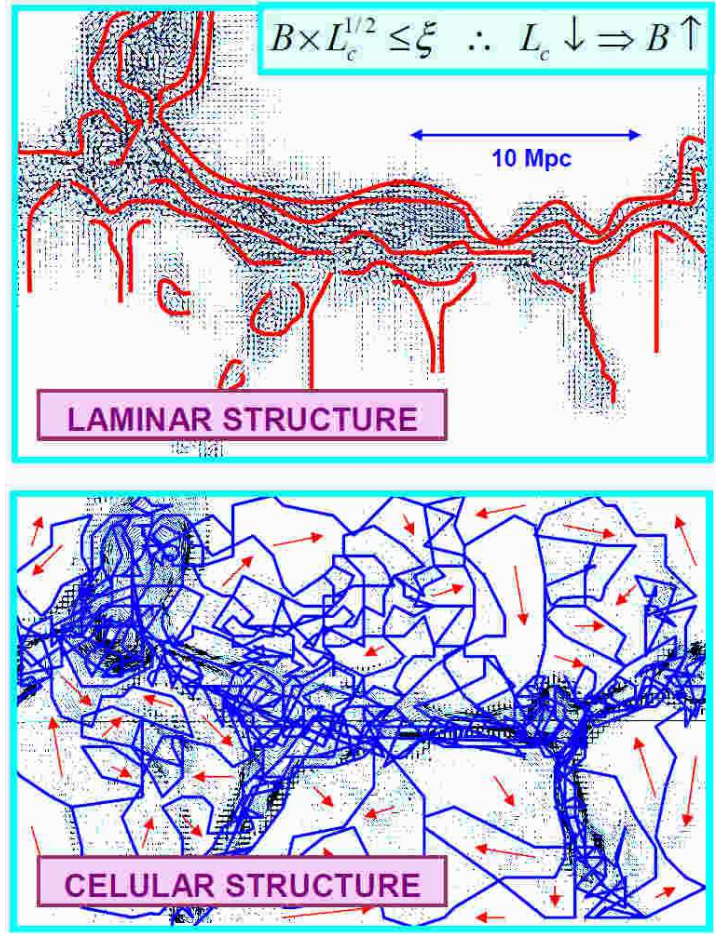}
\end{center}
\caption[]{two possible extreme models for the IGMF structure:
(top) laminar structure and (bottom) cellular structure. These
schematic plots are adaptations by hand made on top of IGMF and
density calculations by \cite{Ryu98}}
\label{fig:IGMFLaminarCelularCartoon}
\end{figure}

Unfortunately, we do not know what is the actual large scale
structure of the IGMF. Nevertheless, we can imagine two extreme
scenarios that are likely to bound the true IGMF structure. In
figure \ref{fig:IGMFLaminarCelularCartoon} calculations of large
scale structure formation by Ryu and co-workers \cite{Ryu98} have
been modified by hand to exemplify these scenarios. The top frame
displays Ryu's IGMF simulation results in the background showing
how, by $z=0$, the magnetic field has been convected together with
the accretion flows into walls, filaments and clusters, depleting
the voids from field. According to these calculations, the
magnetic field is confined in high density, small filling factor
regions, bounded by a rather thin skin of rapidly decreasing
intensity, surrounded by large volumes of negligible IGMF. As
suggested by the free-hand lines on top of the figure, the IGMF
inside structures is highly correlated in scales of up to tens of
Mpc. Furthermore, in order to comply with the rotation measurement
constraints mentioned before, the intensity of the magnetic field
inside the density structures must be correspondingly high,
$0.1$--$1$ $\mu$G, which is comparable with GMF values within the
interstellar medium. We will call the latter scenario {\it
laminar-structure \/}.

The second model, that we will call {\it cellular-structure \/},
is depicted in the bottom panel of figure
\ref{fig:IGMFLaminarCelularCartoon}. We imagine the space divide
into adjacent cells, each one with a uniform magnetic field
randomly oriented. We identify the size of a cell with the
magnitude of the local reversal scale. Furthermore, one can assume
that the intensity of the magnetic field scales as some power of
the local matter (electron) density and, consequently, the
rotation measurement constraint $B_{IGM} \times L_{c}^{1/2} <
10^{-9}$ G $\times$ Mpc$^{1/2}$ tells how the reversal scale,
i.e., the size of the cells, should be scaled. A convenient
reference, such as the IGMF in the Virgo \cite{Arp88} or Comma
\cite{Kim_Coma_IGMF_1989} cluster can be used for normalization.
The cellular-structure scenario leads to a more widespread IGMF,
filling even the voids. The observational constraints imply then
that the IGMF varies much more smoothly, from $~10^{-10}$G inside
voids to a few times $10^{-9}$--$10^{-8}$G inside walls and
filaments, only reaching high values, $~0.1-1$ $\mu$G, inside and
around clusters of galaxies.

\begin{figure}[hb]
\begin{center}
\includegraphics[width=0.8\textwidth]{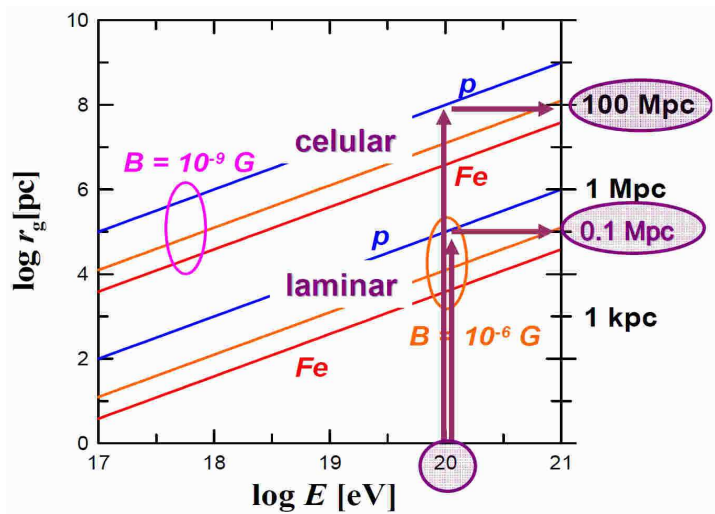}
\end{center}
\caption[]{Typical Larmor radii of nuclei in both IGMF scenarios,
showing the very different scales involved in each model.}
\label{fig:IGMFLaminarCelularRLarmor}
\end{figure}

Observations are not enough at present to distinguish between
these two scenarios. Nevertheless, we can still try to asses what
are their implications for UHECR propagation which, inspecting
figure \ref{fig:IGMFLaminarCelularRLarmor}, must be important.

\subsection{UHECR propagation in a laminar-IGMF} \label{Section:IGMFLaminar}

This is the most difficult scenario to dealt with because it does
not accept a statistical treatment and results are very dependent
on details about the exact magnetic field configuration inside the
GZK-sphere, which is beyond our present knowledge.

\begin{figure}[t]
\begin{center}
\includegraphics[width=0.95\textwidth]{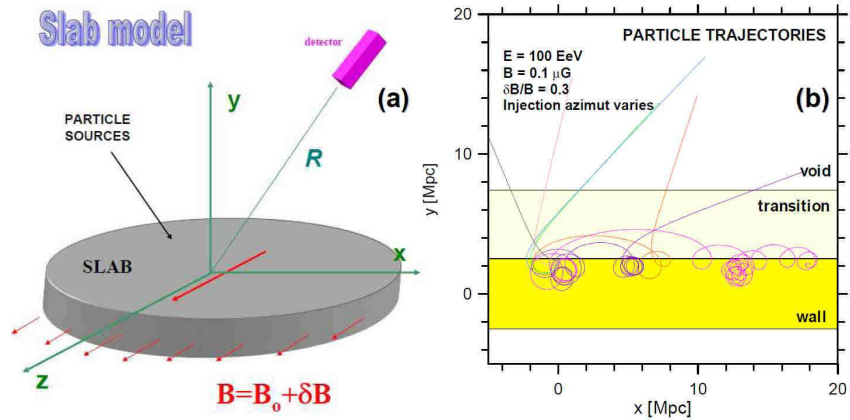}
\end{center}
\caption[]{(a) Simplified model of a wall, or slab, containing
UHECR sources and surrounded by a void. The magnetic field
configuration is representative of the laminar model. (b) Cross
section of the wall in (a) at the plane $z=0$. Several particles
trajectories are shown for proton injection at $E=100$ EeV.
Adapted from \cite{GMT_APJL98_IGMF}.} \label{fig:IGMFSlabModel}
\end{figure}

A simpler approach is to study the UHECR emissivity of a single
wall surrounded by a void \cite{GMT_IGMF_ECRS99,GMT_APJL98_IGMF}.
Figure \ref{fig:IGMFSlabModel}.a shows the corresponding model for
a wall immersed in a void; the magnetic field inside the wall has
two components, o uniform field along the $z$-axis of intensity
$0.1$ $\mu$G and a random component with a Kolmogoroff power law
spectrum of amplitude equal to 30\% of the regular component. One
hundred UHECR sources are included inside the wall, and each one
of them injects protons at the same rate and with the same power
low energy spectrum, $dN_{inj}/dE \propto E^{-2}$. Pair-production
and photo-pion production losses in interactions with the infrared
and microwave backgrounds respectively are also included. The wall
has a radius of $20$ Mpc, a thickness of $5$ Mpc and is sandwiched
by a transition layer $5$ Mpc in thickness where the magnetic
field decreases exponentially up to negligible values inside the
surrounding void. Once the system reaches steady state, a detector
can be shifted around the wall to simulate observers at arbitrary
positions with respect to the wall. In a real situation, this
system could be representative, for example, of the supergalactic
plane; in that case the Milky Way, i.e. we, the observers, should
be located at some point on the $x$-$z$ plane (but we don't know
at what angle with respect to the $z$-axis). The simulations show
that the UHECR flux measured can vary by three orders of magnitude
depending of the relative orientation between the wall, the field
and the observer. At the same time, almost all directional
information is lost, and the strength of the GZK-cut-off would
vary considerably as a function of orientation
\cite{GMT_APJL98_IGMF}.

The previous effects can be intuitively understood by looking at
figure \ref{fig:IGMFSlabModel}.b, which shows a cross section of
the wall in figure \ref{fig:IGMFSlabModel}.a at the plane $z=0$.
Several particles trajectories are shown for proton injection at
$E=100$ EeV, with different azimuthal angles and a slight
elevation with respect to the $x$-$y$ plane. It can be seen that
there is nothing like a random walk: particles tend to be trapped
inside the wall and move in a systematic way. Most of the
particles drift perpendicularly to the regular field while their
guiding centers bounce along the field. It can also be seen how
the gyroradii decrease as the particles lose energy in
interactions with the radiation backgrounds. Even the few
particles that do escape from the wall, do so in an non-isotropic
manner (e.g., predominantly to the right for $y>0$).

The laminar IGMF model is, actually, the worst scenario for doing
some kind of astronomy with UHECR. It would be very difficult to
interpret the UHECR angular data and to identify individual
particle sources. Furthermore, the significance of any statistical
analysis would be greatly impaired due to systematics.

Further studies on this model can also be found, for example, in
references \cite{SiglLemoineBiermann99,LemoineSiglBiermann99}.

\subsection{UHECR propagation in a cellular-IGMF} \label{Section:IGMCellular}

The cellular model is the easiest scenario to deal with
numerically and, by far, the most promising from the point of view
of the astrophysics of UHECR. This is also the IGMF model that has
been used probably more frequently in the literature
\cite{Auger_Design_Report,GMT_Durban_Cluster97,GMT_IGM_nondiffusive_prop,GMT_Durban_IGMF97,GMT_APJL98_cluster,GMT_APJL99_spectrum,UHECRiso00,UHECRghosts01}.

The main assumption here, is that the intensity of the magnetic
field scales with density. Indeed, for those spatial scales where
measurements are available, the intensity of the magnetic fields
seem to correlate remarkably well with the density of thermal gas
in the medium. This is valid at least at galactic and smaller
scales \cite{Vallee97,GMT_Duvernois_2000}. It is apparent that B
can be reasonably well fitted by a single power law over $\sim 14$
orders of magnitude in thermal gas density at sub-galactic scales.
A power law correlation, though with a different power law index,
is also suggested at very large scales (c.f. figure 5 in Medina
Tanco 2000b), from galactic halos to the environments outside
galaxy clusters, over $\sim 4$ orders of magnitude in thermal gas
density. This view \cite{Vallee97} is, however, still
controversial \cite{IGMF_Kronberg}. In fact, magnetic fields in
galaxy clusters are roughly $\sim 1$ $\mu$G, which is of the order
of interstellar magnetic fields; furthermore, supracluster
emission around the Coma cluster suggests $\mu$G fields in
extended regions beyond cluster cores. The latter could indicate
that the IGMF cares little about the density of the associated
thermal gas density, having everywhere an intensity close to the
microwave background-equivalent magnetic field strength, $B_{BGE}
\simeq 3 \times 10^{-6}$ G.

Taking the view that a power law scaling exist, a model can be
devised in which the IGMF correlates with the distribution of
matter as traced, for example, by the distribution of galaxies. A
high degree of non-homogeneity should then be expected, with
relatively high values of $B_{IGMF}$ over small regions ($
\stackrel{~}{<} 1$ Mpc) of high matter density. These systems
should be immersed in vast low density/low $B_{IGMF}$ regions with
$B_{IGMF} < 10^{-9}$ G. Furthermore, in accordance with rotation
measurements, the topology of the field should be such that it is
structured coherently on scales of the order of the correlation
length $L_{c}$ which, in turn, scales with IGMF intensity: $L_{c}
\propto B_{IGMF}^{-2}(r)$. $\vec{B}_{IGMF}$ should be
independently oriented at distances $> L_{c}$. Therefore, a 3D
ensemble of cells can be constructed, with cell size given by the
correlation length, $L_{c}$, and such that: $L_{c} \propto
B_{IGMF}^{-2}(r)$, while $B_{IGMF} \propto \rho_{gal}^{0.35}(r)$
\cite{Vallee97} or $\propto \rho_{gal}^{2/3}(r)$ (for frozen-in
field compression), where $\rho_{gal}$ is the galaxy density, and
the IGMF is uniform inside cells of size $L_{c}$ and randomly
oriented with respect to adjacent cells
\cite{GMT_Durban_IGMF97,UHECRghosts01}. The observed IGMF value at
some given point, like the Virgo cluster, can be used as the
normalization condition for the magnetic field intensity. The
density of galaxies, $\rho_{gal}$, is estimated using either
redshift catalogs [like the CfA Redshift Catalogue
\cite{GMT_APJL98_cluster,GMT_APJL99_spectrum} or the PSCz
\cite{Blanton:Blasi:Olinto:spectrum:2000}], or large scale
structure formation simulations \cite{UHECRghosts01}. The latter
is a convenient way to cope with, or at least to assess the
importance of, the several biases involved in the use of galaxy
redshift surveys to sample the true spatial distribution of matter
in 3D space.

The relevant energy losses for UHECR during propagation are: pair
production via $\gamma$--$\gamma$ with CMB for photons, redshift,
pair production and photopion production in interactions with the
CMB for nuclei and, for heavy nuclei, also photo-disintegration in
interactions with the infrared background. All of these can be
appropriately included
\cite{Berezinsky88,photo_pion_Yoshida_Teshima_1993,Stecker_photodisintegration_1999,TopDownRevSigl}.

The spatial distribution of the sources of UHECR is tightly linked
to the nature of the main particle acceleration/production
mechanism involved. However, in most models, particles will either
be accelerated at astrophysical sites that are related to baryonic
matter, or produced via decay of dark matter particles. In both
cases the distribution of galaxies (luminous matter) should be an
acceptable, if certainly not optimal, tracer of the sources.

\begin{figure}[t]
\begin{center}
\includegraphics[width=0.9\textwidth]{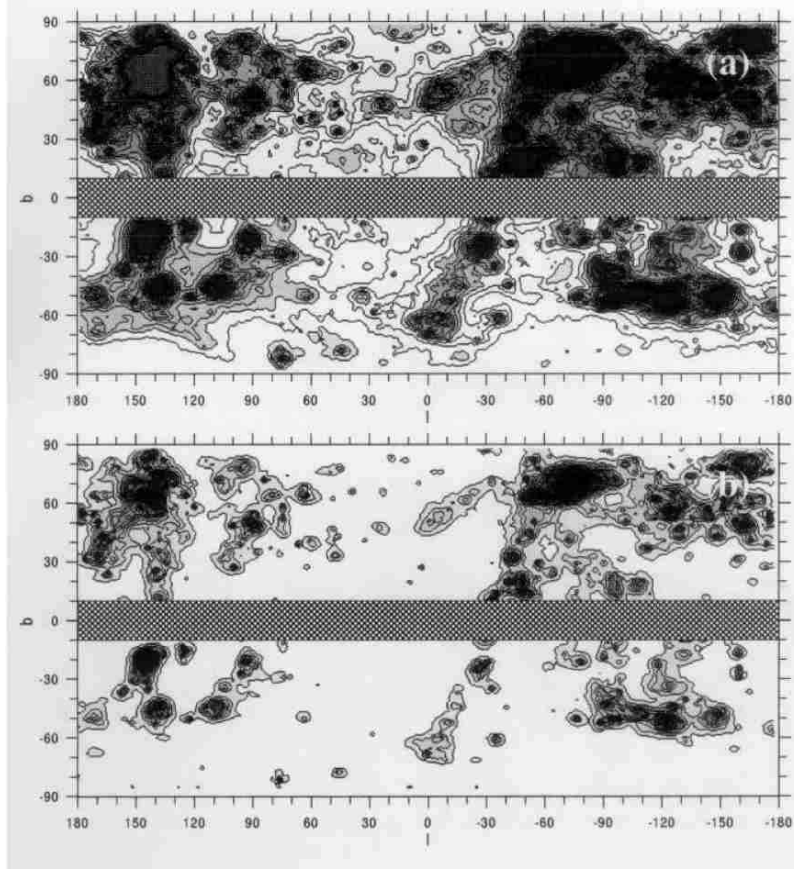}
\end{center}
\caption[]{Arrival probability distribution of protons (linear
scale) as a function of galactic coordinates for a distribution of
sources following the distribution of luminous matter inside $100$
Mpc. $dN_{inj}/dE \propto E^{-2}$, with (a) $E_{inj} > 4 \times
10^{19}$ eV and (b) $E_{inj}> 10^{20}$ eV.}
\label{fig:IGMFArrivalProbabDistCelular}
\end{figure}

Once the previously described scenario is built, test particles
can be injected at the sources and propagated through the
intergalactic medium and intervening IGMF to the detector at
Earth.

Figure \ref{fig:IGMFArrivalProbabDistCelular}a-b show the arrival
probability distribution of UHECR protons as a function of
galactic coordinates for a distribution of sources following the
distribution of luminous matter inside $100$ Mpc(CfA2 catalog). A
power law injection energy spectrum at the sources is assumed,
$dN_{inj}/dE \propto E^{-2}$, with (a) $E_{inj} > 4 \times
10^{19}$ eV and (b) $E_{inj}> 10^{20}$ eV respectively.

\begin{figure}[t]
\begin{center}
\includegraphics[width=0.8\textwidth]{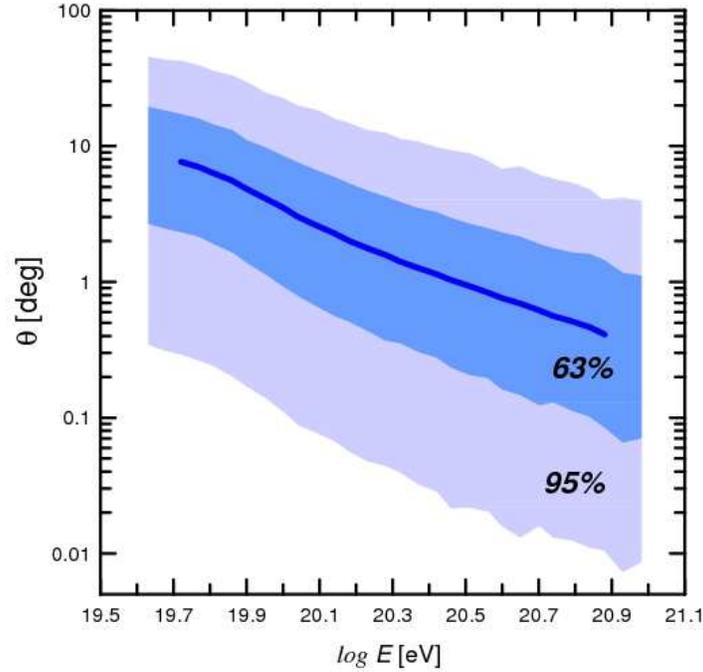}
\end{center}
\caption[]{Median and 63\% and 95\% C.L. for the deflection angle
of an incoming UHECR proton with respect to the true angular
position of the source for the example in figure
\ref{fig:IGMFArrivalProbabDistCelular}. All sky average.}
\label{fig:IGMFSlabDeflection}
\end{figure}

It can be seen that, in contrast to the laminar IGMF case, in this
scenario information regarding the large scale distribution of the
sources inside the GZK-sphere can be easily recoverable. The
supergalactic plane and the Virgo cluster, in particular, are
clearly visible between $l\simeq 0$--$-100$. It can also be
appreciated the increase in resolution as the energy reaches the
$100$ EeV range and the gyroradii of UHECR protons become
comparable to the size of the GZK-sphere. It is also in the
cellular model that the deflection angle of the incoming particle
with respect to the true angular position of the source (see
figure \ref{fig:IGMFSlabDeflection}) is small enough for an UHECR
astronomy to develop at the largest energies.

\subsection{Anisotropy observations and magnetic fields} \label{Section:AnisoObservationsMF}

CR anisotropies can be divided, in principle, according to the
angular scales they affect at a given energy region: (a) large
scales, extending across several tens of degrees in the sky, (b)
medium scales, $\lesssim 10^o$ and (c) small scales, of the order
of the angular resolution of the experiment, i.e, $\sim 1-3^o$.

At all energies CR are remarkably isotropic. At energies below
$\sim 100$ TeV, a first harmonic analysis shows an anisotropy
amounting to less than 0.07 \%. At large energies, on the other
hand, measurements are increasingly difficult due to lowering
statistics and ambiguities in the interpretation of the data due
to the non-uniformity of the detectors´ acceptance
\cite{AnisoWatsonLowEnergy84,AnisoWatsonHighEnergy91}.
Nevertheless, all the data available is consistent with large
scale isotropy
\cite{AGASAIsotropy1,AGASAIsotropy2,HiResIsotropy1,AugerICRCMantsch}.

In fact, besides the AGASA experiment, neither HiRes or Auger have
been able so far of detecting anisotropy at any energy or angular
scale
\cite{HiResIsotropy1,AugerICRCMantsch,AugerICRCAnisoPrescription}.

At energies around 1 EeV, i.e., the beginning of the ankle where
there should still be a sizable Galactic contribution to the flux,
Fly´s Eye has encountered a but statistically significant
correlation with the Galactic plane in the energy range between
0.2 and 3.2 EeV \cite{FlysEyeAnisoGalPlane}. They assessed the
probability of this result being a statistical fluctuation of an
isotropic distribution at $< 0.06$\%. The most significant
enhancement is in the interval 0.4-1.0 EeV. AGASA, on the other
hand has postulated a $>4 \sigma$ excess towards the direction of
the Galactic center \cite{AGASAAniso1EeVGalCenter}. The excess was
confirmed in two independent data sets with 18274 and 10933 events
in the 1-2 EeV region, with chance probabilities of 0.3 and 0.5\%
respectively. The $4.5 ~\sigma$ effect observed corresponds to 506
events detected in a region of the sky where only 413.6 were
expected. Associated with this enhancement was a probably dipolar
signal towards the inner regions of the Galaxy of amplitude 0.04
\cite{AGASAAniso1EeVGalCenterICRC}. An independent confirmation of
an anisotropy likely related with this one comes from the SUGAR
experiment \cite{SUGARAniso1EeVGalCenter} which, different from
AGASA, was able to look directly at the Galactic center.
Unfortunately, these findings have not been confirmed so far by
either HiRes or Auger \cite{AugerAniso1EeVICRC}.

At small scales, $< 2.5^o$, comparable with the resolution of the
experiment, AGASA has postulated the existence of pairs and
triplets of events, which have grown in number over the years to a
total of 7 pairs and 1 triplet (or 9 pairs if the triplets is
counted as 2 pairs) above $40$ EeV
\cite{AGASApairs1,AGASAPairs2,AGASAPairs3,AGASAPairs4}. Since,
following AGASA estimations, a total of 1.7 pairs was expected at
this separation, the results has a chance probability of less than
0.1\%. This result has not been confirmed by HiRes in the combined
AGASA-HiRes data set \cite{HiResAGASAClustering}, but still
remains a topic of hot debate due to its enormous astrophysical
significance. It must also be noted that it is difficult to
understand simultaneously the existence of at least the three
original pairs when the actual distribution of matter inside the
GZK sphere is taken into account \cite{GMT_APJL98_cluster}.

Finally, another very promising anisotropic signal coming from the
AGASA experiment is found as an alignment in the relative
orientation of pairs of incoming events above 10 EeV in the
$\Delta l$--$\Delta b$ plane (Galactic coordinates) on scales of
$\lesssim 10^o$
\cite{AGASAPolarizationHamburg,AGASAPolarizationTsukuba}. It must
be noted that this anisotropy is fundamentally different from a
simple clustering of events in a given angular scale, since it is
limited to an aligned structure in the two point correlation
function. This signal can only be produced as the result of
charged CR bending their trajectories in the Galactic magnetic
field. The astrophysical implications of this observation have
been analyzed in detail in \cite{AGASAPolarizationGMT}. CR
polarization, if confirmed, could turn into a powerful tool for
the determination of the number of nearby CR point sources and for
imposing constraints on the intensity and topology of the Galactic
and extragalactic magnetic fields.

Summing up the results on anisotropy at the highest energies so
far, and remembering our discussion about the effects of different
spacial structures and intensities of the IGMF (sections
\ref{Section:IGMFLaminar} and \ref{Section:IGMCellular}), the high
degree of isotropy observed so far by most experiments, seems to
favor a laminar IGMF structure. However, it must be remembered
that local coherent magnetized structures, as our own Halo could
be, may be able to de-focus particles coming from point sources
into an apparently isotropic flux
\cite{MagnetizedHaloM87a,MagnetizedHaloM87b} further complicating
the analysis.


\printindex
\end{document}